\documentclass[a4paper, oneside, twocolumn, notitlepage, 10pt]{extarticle_ecoc}
\usepackage{ecoc}

\begin{document}
\selectlanguage{english}    


\title{\!\!Learning for Perturbation-Based Fiber Nonlinearity Compensation\!\!}%


\author{
    Shenghang~Luo\textsuperscript{(1)}, Sunish~Kumar~Orappanpara~Soman\textsuperscript{(2)},
    Lutz~Lampe\textsuperscript{(1)}, Jeebak~Mitra\textsuperscript{(3)}, \\ and Chuandong~Li\textsuperscript{(3)}
}

\maketitle                  


\begin{strip}
 \begin{author_descr}

   \textsuperscript{(1)} Department of Electrical and Computer Engineering, The University of British Columbia, Vancouver, BC V6T 1Z4, Canada,
   \textcolor{blue}{\uline{shenghang@ece.ubc.ca}}, \textcolor{blue}{\uline{lampe@ece.ubc.ca}}  

   \textsuperscript{(2)} School of Engineering, Ulster University, Newtownabbey, BT37 0QB, United Kingdom,
   \textcolor{blue}{\uline{s.orappanpara\_soman@ulster.ac.uk }} 

   \textsuperscript{(3)} Huawei Technologies Canada, Ottawa, ON K2K 3J1, Canada,
   \textcolor{blue}{\uline{jeebak.mitra@huawei.com}}, \textcolor{blue}{\uline{chuandong.li@huawei.com}}

 \end{author_descr}
\end{strip}

\setstretch{1.1}
\renewcommand\footnotemark{}
\renewcommand\footnoterule{}


\begin{strip}
  \begin{ecoc_abstract}
    Several machine learning inspired methods for perturbation-based fiber nonlinearity (PB-NLC) compensation have been presented in recent literature. We critically revisit acclaimed benefits of those over non-learned methods. Numerical results suggest that learned linear processing of perturbation triplets of PB-NLC is preferable over feedforward neural-network solutions.  
\textcopyright2022 The Author(s)
  \end{ecoc_abstract}
\end{strip}


\section{Introduction}
The Kerr-induced fiber nonlinearity puts a cap on the maximum achievable information capacity in long-haul coherent optical transmission systems \cite{9,19,20}. Over the last few years, the advances in deep learning algorithms initiated the development of several system-agnostic  machine learning (ML) approaches to compensate for the fiber Kerr nonlinearity effects. On this ground, learned versions of the perturbation theory-based fiber nonlinearity compensation (PB-NLC) technique have  widely been investigated and demonstrated its effectiveness in estimating the complex nonlinear distortion field with the perturbation triplets as the input features \cite{5,2,6,7,8,10,4}. 

Since the overall computational complexity of the PB-NLC techniques has often been  considered a limitation for practical implementation, one direction of learned PB-NLC has been to lower the complexity by, for example,  reducing the input feature vector size and pruning for neural network (NN) based PB-NLC \cite{6,10,2}. On the other hand, the conventional (CONV) PB-NLC technique has also notably been optimized in terms of nonlinearity compensation capability and overall computational complexity. Nonlinearity compensation performance improvements have been achieved by using realistic pulse shapes and the inclusion of the power profile in the coefficient computation \cite{12,13}. Significant complexity reductions have been obtained by coefficient quantization \cite{3}. Furthermore, the introduction of a cyclic buffer (CB) in the triplet computation stage permits further complexity savings for both learned and non-learned PB-NLC \cite{4}.



In this paper, we revisit the acclaimed performance  benefits  of learned PB-NLC approaches\cite{5,2} over their non-learned counterparts. For this, we carefully evaluate the overall computational complexity of existing learned and non-learned PB-NLC techniques, considering the available advancements for all of them. We note that such a comparison has not been done in the references proposing learned PB-NLC. As one important finding, our results show that feedforward NN (FNN)-based PB-NLC has hardly advantages over non-learned PB-NLC. On the other hand, the approach of learning PB-NLC coefficients from a least-squares (LS) optimization is shown to outperform the best non-learned PB-NLC variant and to  
provide the best performance-complexity trade-off of all tested methods. 

\begin{table*}[!t]
   \centering
\caption{List of Techniques}
\label{tab:table1}%
      \begin{tabular}{|c|c|c|c|c|c|c|c|c|c|}
         \hline 
         Technique  &  
         Learned/Non-learned  &  
         Complexity reduction method  &  
         Referenced paper \\
         \hline\hline  CONV PB-NLC  & Non-learned & Rounding-based quantization & \cite{14} \\ 
         \hline
        CONV-AM PB-NLC & Non-learned & None & \cite{1} \\
        \hline
        FNN PB-NLC & Learned & Pruning & \cite{2,5,6,7,8}\\
        \hline
        FNN-AM PB-NLC & Learned & Pruning  & Our extension \\
        \hline
         RNN PB-NLC & Learned & None & \cite{10}  \\
        \hline
        CNN PB-NLC & Learned & None & \cite{11}  \\
         \hline
         LS PB-NLC & Learned & CB, clustering-based quantization & \cite{4}  \\
        \hline
      \end{tabular}
\end{table*}

\begin{figure*}[!t]
   \centering
    \includegraphics[width=0.9\linewidth]{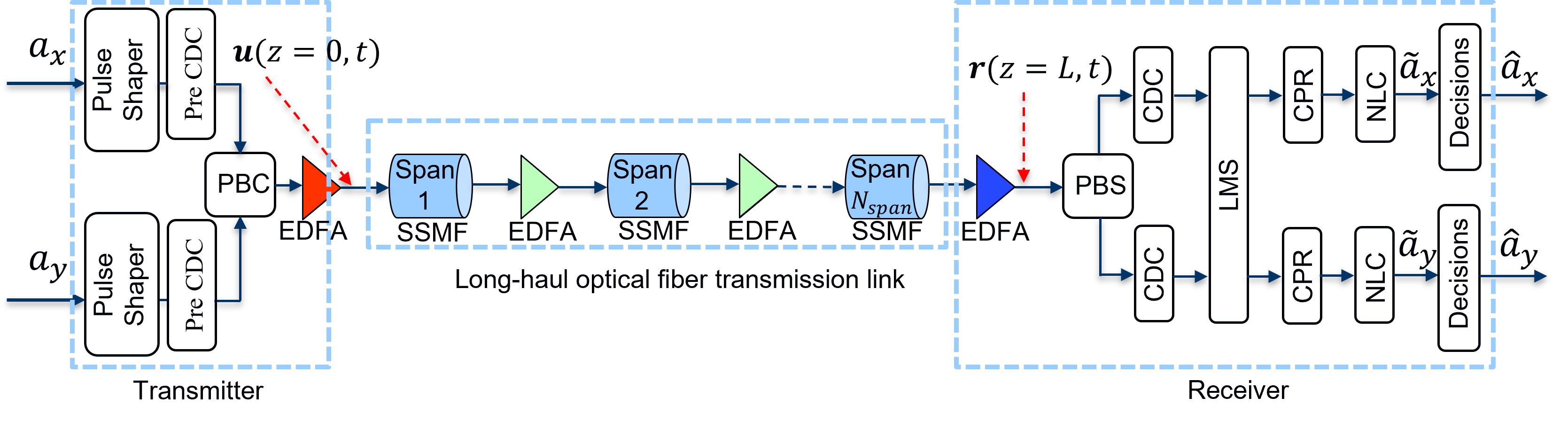}
    \caption{The system model for the WDM system. (PBC: polarization beam combiner, PBS: polarization beam splitter) 
    }
    \label{fig:figure1}
\end{figure*}

\section{PB-NLC Techniques}

In CONV PB-NLC, the nonlinear distortion field is numerically calculated and subtracted from the symbol of interest. 
However, the truncation of the perturbation approximation at first-order in PB-NLC leads to a power overestimation problem and degrades the NLC performance. This can be overcome using an additive-multiplicative PB-NLC (CONV-AM PB-NLC) technique \cite{1}. 

ML-based solutions utilizing the existing PB-NLC distortion model have been investigated in several literatures. In \cite{5,2}, fully connected FNNs with perturbation triplets as the input features have been proposed to estimate the nonlinear distortion field instead of numerically computing it. This technique is referred to as FNN PB-NLC. 
 NNs can also be used to predict the distortions in the above-mentioned additive-multiplicative model, which we refer to as FNN-AM PB-NLC, and for which we show results for the first time in this paper.
 Rather than using an FNN, the application of a recurrent  (RNN) and a convolution neural network (CNN) have been proposed in \cite{10} and \cite{11}, respectively. The RNN solution from \cite{10} has a similar performance as a fully connected FNN but affords a complexity reduction of about 47\%. The CNN solution provides modest complexity benefits when compared with a fully connected FNN in \cite{11}. However, the complexity figures in \cite{11} are extremely high and suggest that an oversized FNN has been considered.  

Besides NN-based PB-NLC, a learned model using the LS method to estimate the perturbation coefficients, referred to as LS PB-NLC, has also been considered in the literature \cite{3}. A summary of the non-learned and learned PB-NLC techniques is provided in Table \ref{tab:table1}. We also state the  complexity-reduction methods that have been applied in the corresponding literature references. 

In this paper, we carry out numerical simulations to compare learned and non-learned PB-NLC techniques. For the former, we focus on FNN PB-NLC, as it has been pitched in \cite{5,2} as superior over CONV PB-NLC, which has motivated this work. We further consider LS PB-NLC, which is closely related to CONV PB-NLC in that it linearly processes perturbation triplets.

\section{System Model}



Fig.~\ref{fig:figure1} shows the simulation setup. We simulate a polarization-multiplexed 5-channel wavelength-division multiplexed (WDM) transmission at 32~Gbaud per channel. The modulation format considered is 16-quadrature amplitude modulation (QAM). The transmitter processing consists of a root-raised cosine (RRC) filter and a 50\% pre-chromatic dispersion compensation (CDC) \cite{6,2}. The signal is then transmitted through 10 spans of a standard single-mode fiber (SSMF) with a span length of 100~km, a dispersion parameter of 17~ps.nm$^{-1}$.km$^{-1}$, a nonlinear parameter of 1.2~W$^{-1}$.km$^{-1}$, an attenuation coefficient of 0.2~dB.km$^{-1}$ and a polarization mode dispersion (PMD) parameter of 0.1~ps.km$^{-\frac{1}{2}}$. An erbium-doped fiber amplifier (EDFA) with a 6~dB noise figure is used to compensate for the span loss. The laser linewidth is 100~kHz.
The receiver processing consists of a 50\% post-CDC, an RRC matched filter, a least mean square (LMS) adaptive 2x2 filter to compensate for PMD, and blind phase search (BPS) for carrier phase recovery (CPR). Then, one of the considered NLC techniques is applied to compensate for the remaining nonlinearity. Following that, the compensated symbols are demodulated to calculate the bit-error rate and Q-factor.


\begin{figure}[t!]
	\hspace*{-5mm}\includegraphics[width=0.55\textwidth]{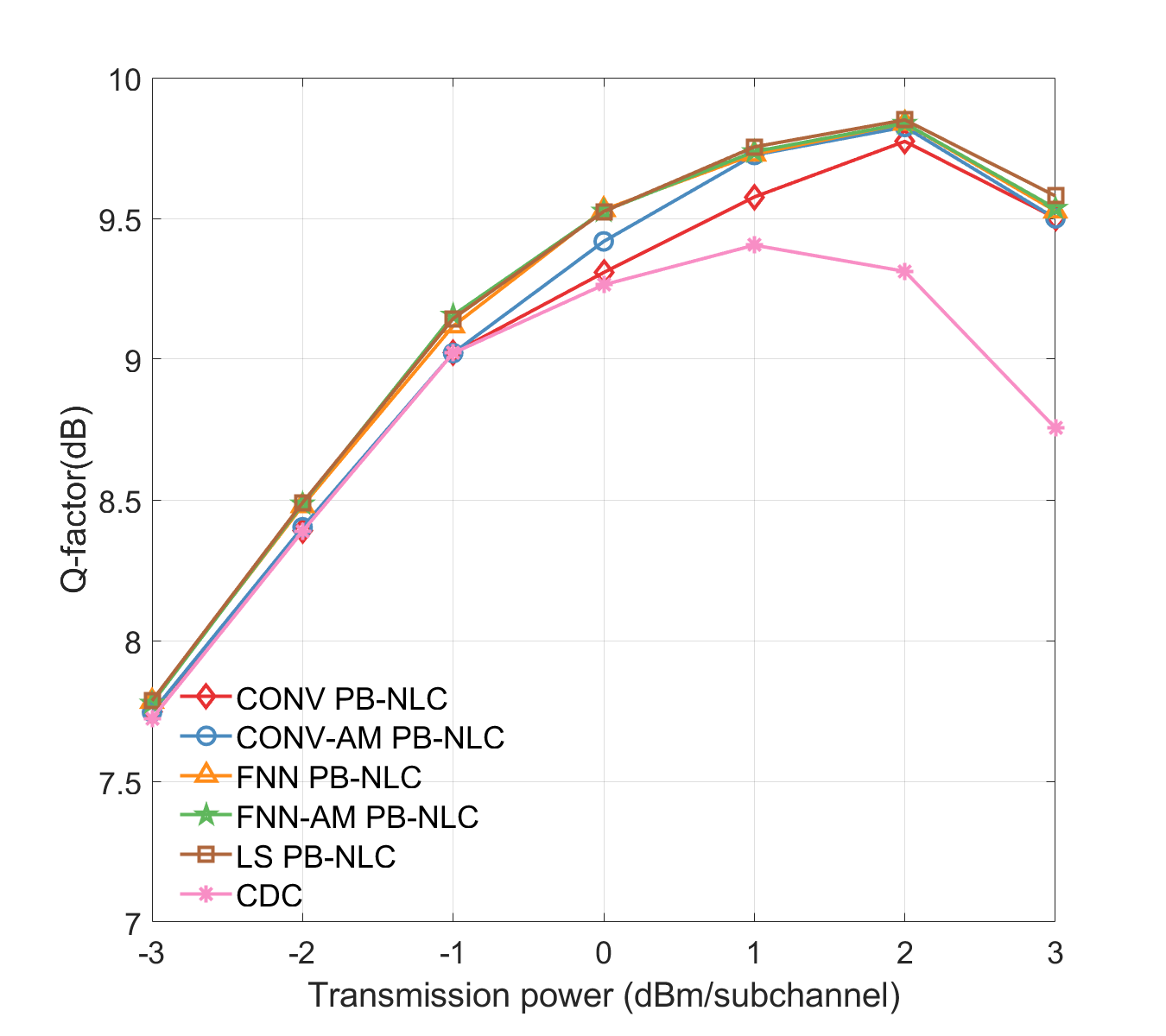}\caption{{Q-factor vs. transmission power.}}
	\label{fig:figure2}
\end{figure}



\section{Results and Discussion}
In the simulation, the center WDM channel is chosen as the target channel to evaluate the performance. Fig.~\ref{fig:figure2} shows the Q-factor performance for all the considered techniques as well as for CDC-only processing for a comparison. We start with a sufficiently large number of triplet features to assess the maximal performance. In particular, the window size selected for the triplet computation is 75 and the number of triplets obtained after applying the truncation criterion in \cite{2} is 1681. We observe that all learned PB-NLC methods perform fairly similar and achieve essentially the same optimal Q-factor at a transmission power of 2~dBm per channel. Furthermore, the CONV-AM PB-NLC reaches an almost identical optimal Q-factor. Hence, without  restrictions on complexity, there is little benefit to the learned approaches. The gain over CDC is about 0.35~dB to 0.45~dB. This may seem moderate when compared to results presented in other literature, but we note that (i) cross-channel NLC due to the WDM transmission limits the achievable gains and (ii) the presence of CPR  partially mitigates the phase noise part of
intra-channel NLC also for CDC-only processing.



\begin{figure}[t!]
	\hspace*{-5mm}\includegraphics[width=0.56\textwidth]{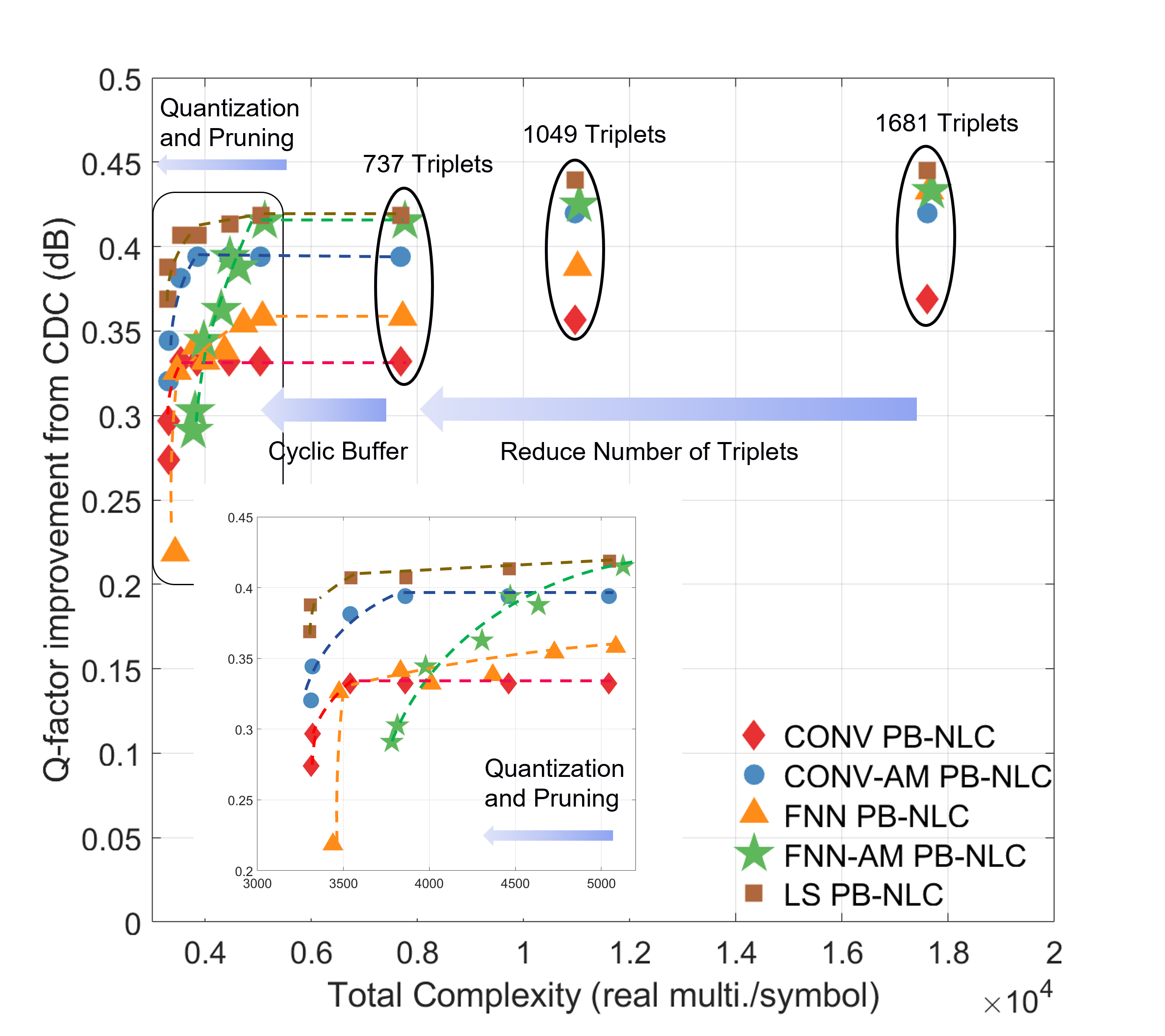}\caption{{Performance vs.\ computational complexity.}}
	\label{fig:figure4}
\end{figure}

Next, we evaluate the Q-factor performance  when reducing the computational complexity for different PB-NLC techniques. 
Fig.~\ref{fig:figure4} shows the Q-factor improvement over CDC as a function of the computational complexity measured as the number of real-valued multiplications per symbol. The complexity reduction is achieved first through smaller symbol-window sizes, which when shortened from 75 to 37 reduces the number of triplets from 1681 to 737. 
Since the performances of CONV-AM PB-NLC, FNN-AM PB-NLC and LS PB-NLC remain close to the best possible values with just an onset of a degradation, we fix 737 triplets and next further reduce complexity by making use of the CB concept \cite{4}. This computationally efficient way of generating triplets does not degrade performance, as seen in Fig.~\ref{fig:figure4}. 

Finally, the number of computations required for processing triplets is lowered through weight-pruning for the NNs and coefficient quantization using K-means clustering for the other methods, respectively.  We use a pruning approach similar to \cite{2}, but rather than applying an absolute threshold, we gradually prune the relatively small weights and rewind the learning rate schedule before fine tuning \cite{LTH}.
The result details are highlighted in the  inset in Fig.~\ref{fig:figure4}.
We observe that LS PB-NLC achieves the best performance-complexity trade-off, followed by CONV-AM PB-NLC. This suggests that PB-NLC based on models that process triplets linearly together with coefficient quantization are more effective than nonlinear processing with FNNs and weight pruning. The total complexity can be reduced to about 3,500 real-valued multiplications per symbol with negligible performance loss, which is more moderate than figures reported in e.g.\ \cite{11} and also competitive with non-PB-NLC-based learned NLC, e.g.\ \cite{co-lstm}.

\section{Conclusions}
In this paper, we have compared conventional and learned PB-NLC techniques  in a 5-channel WDM transmission over a 1,000~km link. Our results suggest that there is little benefit to FNN-based learned PB-NLC. On the other hand, linear triplet processing with coefficients learned from LS optimization is highly effective. Our work highlights the importance of applying available performance improvements (e.g.\ AM PB-NLC) and complexity reduction methods (e.g.\ quantization and CB) when conducting a comparison of learned and non-learned methods. Future work will investigate whether pruned RNN PB-NLC can retain performance better than FNN PB-NLC.




\printbibliography

@ARTICLE{co-lstm,
  author={Ming, Hao and Chen, Xinyu and Fang, Xiansong and Zhang, Lei and Li, Chenjia and Zhang, Fan},
  journal={Journal of Lightwave Technology}, 
  title={Ultralow Complexity Long Short-Term Memory Network for Fiber Nonlinearity Mitigation in Coherent Optical Communication Systems}, 
  year={2022},
  volume={40},
  number={8},
  pages={2427-2434},
  doi={10.1109/JLT.2022.3141404}}

@inproceedings{
LTH,
title={Comparing Rewinding and Fine-tuning in Neural Network Pruning},
author={Alex Renda and Jonathan Frankle and Michael Carbin},
booktitle={International Conference on Learning Representations},
year={2020},
url={https://openreview.net/forum?id=S1gSj0NKvB}
}

@online{1,
  author        = "Liang Xiaojun",
  title         = "Analysis and Compensation of Nonlinear Impairments in Fiber-Optic Communication Systems",
  year          = "2015",
  url           = "https://macsphere.mcmaster.ca/bitstream/11375/17193/2/main_thesis.pdf"
}

@article{2,
  author        = "Zhang, Shaoliang and Yaman, Fatih and Nakamura, Kohei and Inoue, Takanori and Kamalov, Valey and Jovanovski, Ljupcho and Vusirikala, Vijay and Mateo, Eduardo and Inada, Yoshihisa and Wang, Ting",
  title         = "Field and lab experimental demonstration of nonlinear impairment compensation using neural networks",
  journal       = "Nature Communications",
  volume        = "10",
  number        = "3033",
  year          = "2019",
  doi           = "https://doi.org/10.1038/s41467-019-10911-9"
}

@article{3,
  author        = "M. Malekiha and D. V. Plant",
  title         = "Adaptive Optimization of Quantized Perturbation Coefficients for Fiber Nonlinearity Compensation",
  journal       = "IEEE Photonics Journal",
  volume        = "8",
  number        = "3",
  pages         = "17",
  year          = "2016",
  doi           = "10.1109/JPHOT.2016.2566341"
}

@article{4,
  author        = "A. Redyuk and E. Averyanov, O. and Sidelnikov, M. Fedoruk and S. Turitsyn",
  title         = "Compensation of Nonlinear Impairments Using Inverse Perturbation Theory With Reduced Complexity",
  journal       = "Journal of Lightwave Technology",
  volume        = "38",
  number        = "6",
  pages         = "1250-1257",
  year          = "2020",
  doi           = "10.1109/JLT.2020.2971768"
}

@inproceedings{5,
  author        = "V. {Kamalov et al.}",
  title         = "Evolution from {8QAM} live traffic to PS {64-QAM} with Neural-Network Based Nonlinearity Compensation on 11000 km Open Subsea Cable",
  booktitle     = "Optical Fiber Communications Conference and Exposition (OFC)",
  pages         = "1-3",
  year          = "2018"
}

@inproceedings{6,
  author        = "Y. {Gao et al.}",
  title         = "Reduced Complexity Nonlinearity Compensation via Principal Component Analysis and Deep Neural Networks",
  booktitle     = "Optical Fiber Communications Conference and Exposition (OFC)",
  pages         = "1-3",
  year          = "2019"
}

@inproceedings{7,
  author        = "E. F. Mateo and F. Yaman",
  title         = "Nonlinearity Compensation in Modern Submarine Networks",
  booktitle     = "OptoElectronics and Communications Conference (OECC) and International Conference on Photonics in Switching and Computing (PSC)",
  pages         = "1-3",
  year          = "2019",
  doi           = "10.23919/PS.2019.8817736"
}

@inproceedings{8,
  author        = "C. {Huang et al.}",
  title         = "Demonstration of Photonic Neural Network for Fiber Nonlinearity Compensation in Long-Haul Transmission Systems",
  booktitle     = "Optical Fiber Communications Conference and Exhibition (OFC)",
  pages         = "1-3",
  year          = "2020"
}

@article{9,
  author        = "Mitra, Partha P. and Stark, Jason B.",
  title         = "Nonlinear limits to the information capacity of optical fibre communications",
  journal       = "Nature",
  volume        = "411",
  year          = "2001",
  doi           = "https://doi.org/10.1038/35082518"
}

@article{10,
  author        = "Y. {Zhao et al.}",
  title         = "Low-Complexity Fiber Nonlinearity Impairments Compensation Enabled by Simple Recurrent Neural Network With Time Memory",
  journal       = "IEEE Access",
  volume        = "8",
  pages         = "160995-161004",
  year          = "2020",
  doi           = "10.1109/ACCESS.2020.3021146"
}

@article{11,
  author        = "C. {Li et al.}",
  title         = "Convolutional Neural Network-Aided {DP-64 QAM} Coherent Optical Communication Systems",
  journal       = "Journal of Lightwave Technology",
  volume        = "40",
  number        = "9",
  pages         = "2880-2889",
  year          = "2022",
  doi           = "10.1109/JLT.2022.3146839"
}

@inproceedings{12,
  author        = "Y. Gao and J. C. Cartledge and A. S. Karar and S. S. Yam",
  title         = "Reducing the complexity of nonlinearity pre-compensation using symmetric {EDC} and pulse shaping",
  booktitle     = "European Conference and Exhibition on Optical Communication (ECOC 2013)",
  pages         = "1-3",
  year          = "2013",
  doi           = "10.1049/cp.2013.1692"
}

@article{13,
  author        = "Z. Tao and Y. Zhao and Y. Fan and L. Dou and T. Hoshida and J. C. Rasmussen",
  title         = "Analytical Intrachannel Nonlinear Models to Predict the Nonlinear Noise Waveform",
  journal       = "Journal of Lightwave Technology",
  volume        = "33",
  number        = "10",
  pages         = "2111-2119",
  year          = "2015",
  doi           = "10.1109/JLT.2014.2364848"
}

@inproceedings{14,
  title         = "Complexity-reduced digital nonlinear compensation for coherent optical system",
  author        = "Zhenning Tao and Liang Dou and Weizhen Yan and Yangyang Fan and Lei Li and Sho-ichiro Oda and Yuichi Akiyama and Hisao Nakashima and Takeshi Hoshida and Jens C. Rasmussen",
  booktitle     = "Photonics West - Optoelectronic Materials and Devices",
  year          = "2013"
}

@article{19,
  author        = "A. Amari and O. A. Dobre and R. Venkatesan and O. S. S. Kumar and P. Ciblat and Y. Jaou{\"e}n",
  title         = "A Survey on Fiber Nonlinearity Compensation for {400 Gb/s} and Beyond Optical Communication Systems",
  journal       = "IEEE Communications Surveys \& Tutorials",
  volume        = "19",
  number        = "4",
  pages         = "3097-3113",
  year          = "2017",
  doi           = "10.1109/COMST.2017.2719958"
}

@article{20,
  author        = "O. S. S. Kumar and A. Amari and O. A. Dobre and R. Venkatesan",
  title         = "Enhanced Regular Perturbation-Based Nonlinearity Compensation Technique for Optical Transmission Systems",
  journal       = "IEEE Photonics Journal",
  volume        = "11",
  number        = "4",
  pages         = "1-12",
  year          = "2019",
  doi           = "10.1109/JPHOT.2019.2923568"
}

\vspace{-4mm}

\end{document}